\documentclass[english]{article}
\usepackage[T1]{fontenc}
\usepackage[latin9]{inputenc}
\usepackage{amstext}
\usepackage{amssymb}
\usepackage{stmaryrd}

\makeatletter
\newcommand{\lyxaddress}[1]{
	\par {\raggedright #1
	\vspace{1.4em}
	\noindent\par}
}

\@ifundefined{date}{}{\date{}}
\usepackage{cite}

\makeatother

\usepackage{babel}
\begin{document}
\title{Contextuality and Informational Redundancy}
\author{Ehtibar Dzhafarov\textsuperscript{1} and Janne V. Kujala\textsuperscript{2}}
\maketitle

\lyxaddress{\begin{center}
\textsuperscript{1}Purdue University, USA, ehtibar@purdue.edu\\
 \textsuperscript{2}University of Turku, Finland, jvk@iki.fi
\par\end{center}}
\begin{abstract}
A noncontextual system of random variables may become contextual if
one adds to it a set of new variables, even if each of them is obtained
by the same context-wise function of the old variables. This fact
follows from the definition of contextuality, and its demonstration
is trivial for inconsistently connected systems (i.e. systems with
disturbance). However, it also holds for consistently connected (and
even strongly consistently connected) systems, provided one acknowledges
that if a given property was not measured in a given context, this
information can be used in defining functions among the random variables.
Moreover, every inconsistently connected system can be presented as
a (strongly) consistently connected system with essentially the same
contextuality characteristics. \medskip{}

\end{abstract}

\section{Introduction}

Throughout this note we use the standard language and notation of
the Contextuality-by-Default theory (CbD) \cite{DzhCerKuj2017}. Familiarity
with CbD therefore is desirable, although we will provide the meaning
of all the terms and notions we use. As a throughout example to illustrate
the issues, consider a rank-3 cyclic system of random variables \cite{DKC2020},
\begin{equation}
\begin{array}{|c|c|c||c|}
\hline R_{1}^{1} & R_{2}^{1} &  & c=1\\
\hline  & R_{2}^{2} & R_{3}^{2} & c=2\\
\hline R_{1}^{3} &  & R_{3}^{3} & c=3\\
\hline\hline q=1 & q=2 & q=3 & 
\\\hline \end{array}\:,\label{eq:cyclic3}
\end{equation}
where $R_{q}^{c}$ is the random variable representing the outcome
of measuring content $q$ (that which is being measured) in context
$c$ (conditions under which it is measured). The random variables
within each context (e.g., $R_{1}^{1},R_{2}^{1}$) form a \emph{jointly
distributed bunch}, while any two random variables in two different
contexts (e.g., $R_{1}^{1},R_{1}^{3}$ or $R_{1}^{1},R_{3}^{2}$)
are \emph{stochastically unrelated} (possess no joint distribution).
The system is \emph{consistently connected} (has no disturbance) if
the distribution of the variables sharing a content is always the
same. In CbD, systems are generally\emph{ inconsistently connected}.

The immediate motivation for this paper is the appearance of an abstract
\cite{Tezzinetal.} claiming that the extension of the notion of contextuality
to inconsistently connected systems, such as offered by CbD, is impossible,
because it contradicts the conjunction of certain ``core principles.''
Some of these principles are satisfied in CbD trivially. One of them,
however, related to creating new random variables as functions of
the old ones, is known not to hold. The reasons for this are well-understood
and presented in the CbD literature (e.g., in Ref. \cite{DzhCerKuj2017}).
It is not surprising therefore that if one posits this ``principle''
as a principle, it rules out CbD. Moreover, we show in this paper
that, with careful formulations of the concepts involved, the ``principle''
in question also contradicts more traditional approaches to contextuality,
those confined to consistently connected systems. In other words,
the alleged principle is in conflict with the very notion of contextuality,
rather than just CbD or other extensions of contextuality to inconsistently
connected systems. 

The purpose of this paper, however, is not entirely polemical. Our
analysis leads us to a better understanding of the structure of the
systems of random variables, specifically of what information in them
can be accessed and utilized in defining new random variables based
on the old ones. 

\section{Connections and their functions}

A \emph{content-context matrix}, like (\ref{eq:cyclic3}), typically
contains empty cells, or ``non-measurements,'' representing cases
when a given content is not measured in a given context. In CbD, the
empty cells can always be viewed as containing deterministic variables
\cite{Dzh2017Nothing} (those equal to some value with probability
1). For instance, (\ref{eq:cyclic3}) can be presented as 
\begin{equation}
\begin{array}{|c|c|c||c|}
\hline R_{1}^{1} & R_{2}^{1} & \alpha & c=1\\
\hline \beta & R_{2}^{2} & R_{3}^{2} & c=2\\
\hline R_{1}^{3} & \gamma & R_{3}^{3} & c=3\\
\hline\hline q=1 & q=2 & q=3 & 
\\\hline \end{array}\:,\label{eq:CbDdet}
\end{equation}
where $\alpha$ stands for a random variable $R_{3}^{1}$ such that
$p\left[R_{3}^{1}=\alpha\right]=1$ (and analogously for $\beta,\gamma$).
This system is equivalent to (\ref{eq:cyclic3}) in the sense that
their contextuality status and the degree of contextuality in them
are precisely the same. In more traditional approaches to contextuality,
confined to consistently connected systems, filling of the empty cells
with deterministic variables is not allowed. Nevertheless, the empty
cells in any system, even if consistently connected, provide certain
information (``no measurement output in this cell''), and this information
can be utilized when one introduces new variables as functions of
old ones. This simple observation is central for the present paper. 

In CbD, the random variables sharing a content are referred to as
\emph{connections }(between bunches). In Ref. \cite{Tezzinetal.}
connections are called \emph{observables}. The latter term is standard
in quantum mechanics, but its use is not natural in CbD, for four
reasons. \emph{First}, CbD has multidisciplinary applications, and
outside quantum mechanics the term ``observable'' may not be well-defined.
\emph{Second}, and most important, observables are defined by contents
only (by what is being measured), while in CbD contexts play an equally
important role --- which is why the distributions within a connection
are generally different in different contexts. Even in quantum mechanics
the random variables in a connection are generally determined not
only by a quantum observable but also by factors, such as quantum
states and signaling, that depend on contexts. See, for example, the
analysis of ``signaling in time'' in Refs. \cite{Bacciagaluppi(2015),KoflerBrukner2013}
 in sequential quantum measurements (where later measurements can
be affected by settings for the previous ones). . \emph{Third}, unlike
the bunches of random variables within contexts, connections are not
random vectors, as they consist of pairwise stochastically unrelated
variables. \emph{Finally}, as explained in Section \ref{sec:Consistification},
connections in CbD can be redefined in multiple ways without affecting
the contextuality status of a system (i.e., preserving the degree
of contextuality in it, by conventional measures). As a result, one
can have contextually equivalent systems with very different sets
of connections.

We will adopt the notation $\mathcal{R}_{q}$ for the connection corresponding
to content $q$ (e.g., $\mathcal{R}_{1}$ for $q=1$, $\mathcal{R}_{2}$
for $q=2$, etc.). Note that the use of the script letter $\mathcal{R}$
is to indicate that a connection possesses no joint distribution of
its elements. In view of the simple observation above, we will expand
the definition of a connection here to also include non-measurements.
Thus, in our example,
\begin{equation}
\mathcal{R}_{1}=\begin{array}{|c|}
\hline R_{1}^{1}\\
\hline \boxempty\\
\hline R_{1}^{3}
\\\hline \end{array},\mathcal{R}_{2}=\begin{array}{|c|}
\hline R_{2}^{1}\\
\hline R_{2}^{2}\\
\hline \boxempty
\\\hline \end{array},\mathcal{R}_{3}=\begin{array}{|c|}
\hline \boxempty\\
\hline R_{3}^{2}\\
\hline R_{3}^{3}
\\\hline \end{array}\:,
\end{equation}
where $\boxempty$ stands for the dummy variable whose single value
is ``no measurement output.'' We can now define the notion of some
connection being a function of some $k$ connections. Without loss
of generality, let them be $\mathcal{R}_{0}$ and $\left(\mathcal{R}_{1},\ldots\mathcal{R}_{k}\right)$,
respectively. We will say that
\begin{equation}
\mathcal{R}_{0}=f\left(\mathcal{R}_{1},\ldots\mathcal{R}_{k}\right),
\end{equation}
if in every context $c$,
\begin{equation}
R_{0}^{c}=f\left(R_{1}^{c},\ldots,R_{k}^{c}\right),
\end{equation}
with the understanding that if a variable $R_{q}^{c}$ is undefined,
it is replaced with $\boxempty$. In CbD, one can simply view $\boxempty$
as the single value of a deterministic variable, $R_{q}^{c}=\boxempty$,\footnote{\label{fn:We-need-to}We need to add a technical detail here. In CbD,
all variables within a connection should have the same set of possible
values. Therefore, if in $\mathcal{R}_{1}$ the variables $R_{1}^{1}$
and $R_{1}^{3}$ are dichotomic, $\pm1$, then they have to be redefined
as having values $\left\{ 1,-1,\boxempty\right\} $, with the understanding
that $p\left[R_{1}^{1}=\boxempty\right]=p\left[R_{1}^{3}=\boxempty\right]=0$.
The non-measurement variable $R_{1}^{3}$ should also be redefined
as having values $\left\{ 1,-1,\boxempty\right\} $, with $p\left[R_{1}^{2}\not=\boxempty\right]=0$.} but in view of the approaches confined to consistently connected
systems, it is better to treat the random variables in a system and
its $\boxempty$ entries separately. For a given context $c$, let
us agree to write $\mathcal{R}_{q}=a$ if $R_{q}^{c}=a$ or if $a=\boxempty$
in the $\left(q,c\right)$-cell. 

\section{From noncontextual to contextual systems}

Let $X,Y,Z$ be $\pm1$-variables with $p\left[X=1\right]=p\left[Y=1\right]=p\left[Z=1\right]=1/2$,
and let them be pairwise stochastically unrelated. The variant of
the system (\ref{eq:cyclic3}) shown below is consistently connected:
\begin{equation}
\begin{array}{|c|c|c||c|}
\hline X & X & \boxempty & c=1\\
\hline \boxempty & Y & Y & c=2\\
\hline Z & \boxempty & Z & c=3\\
\hline\hline q=1 & q=2 & q=3 & 
\\\hline \end{array}\:.\label{eq:variant1 cyclic3}
\end{equation}
This system is noncontextual because it has a \emph{coupling} with
\emph{(multi)maximal connections}:
\begin{equation}
\begin{array}{|c|c|c||c|}
\hline S & S & \boxempty & c=1\\
\hline \boxempty & S & S & c=2\\
\hline S & \boxempty & S & c=3\\
\hline\hline q=1 & q=2 & q=3 & 
\\\hline \end{array}\:,\label{eq:coupling 0}
\end{equation}
where $S$ is a $\pm1$-variable with $p\left[S=1\right]=1/2$. By
definition, it is a coupling of (\ref{eq:variant1 cyclic3}) because
(a) all variables in it are jointly distributed, and (b) in every
context the joint distribution of the variables is the same as in
system (\ref{eq:variant1 cyclic3}). This coupling has (multi)maximal
connections because in each connection all the random variables are
pairwise equal to each other with maximal possible probability (which
in this example, as in all consistently connected systems, is 1).\footnote{The prefix ``multi'' is to indicate that the maximal probability
is achieved for all pairs of the random variables in a connection,
but since in our example a connection contains only two variables,
the notions of multimaximality and maximality coincide.}

Let us add a new connection $\mathcal{R}_{4}$ to this system: 
\begin{equation}
\begin{array}{|c|c|c|c||c|}
\hline X & X & \boxempty & 1 & c=1\\
\hline \boxempty & Y & Y & \boxempty & c=2\\
\hline Z & \boxempty & Z & -1 & c=3\\
\hline\hline q=1 & q=2 & q=3 & q=4 & 
\\\hline \end{array}\:.\label{eq:variant 1 with addition 1}
\end{equation}
The system remains noncontextual, because, as mentioned earlier with
reference to \cite{Dzh2017Nothing}, contextuality status of a system
does not change if one adds to or deletes from it deterministic variables.
The new system, however, is no longer consistently connected (because
being equal to 1 with probability 1 and being equal to $-1$ with
probability 1 describes two different distributions). Let us further
introduce another connection, $\mathcal{R}_{0}$: 

\begin{equation}
\begin{array}{|c|c|c|c|c||c|}
\hline X & X & X & \boxempty & 1 & c=1\\
\hline \boxempty & \boxempty & Y & Y & \boxempty & c=2\\
\hline -Z & Z & \boxempty & Z & -1 & c=3\\
\hline\hline q=0 & q=1 & q=2 & q=3 & q=4 & 
\\\hline \end{array}\:.\label{eq:contextual 1}
\end{equation}
Observe now that $\mathcal{R}_{0}$ can be obtained as a function
of other connections in the system,
\begin{equation}
\mathcal{R}_{0}=\left\{ \begin{array}{ccc}
\mathcal{R}_{1} & \textnormal{if} & \mathcal{R}_{4}=1\\
\mathcal{-R}_{1} & \textnormal{if} & \mathcal{R}_{4}=-1\\
\boxempty & \textnormal{if} & \textnormal{otherwise}
\end{array}\right\} =f\left(\mathcal{R}_{1},\mathcal{R}_{4}\right).
\end{equation}
Note that ``if'' here means ``in any context c in which.'' As
a function of other connections, therefore, $\mathcal{R}_{0}$ is
\emph{informationally redundant}, and the authors of Ref. \cite{Tezzinetal.}
think this means that the addition of $\mathcal{R}_{0}$ to the noncontextual
system (\ref{eq:variant 1 with addition 1}) should leave the system
noncontextual. However, the new system is contextual. This is easy
to see by considering its subsystem
\begin{equation}
\begin{array}{|c|c|c||c|}
\hline X & X & \boxempty & c=1\\
\hline \boxempty & Y & Y & c=2\\
\hline -Z & \boxempty & Z & c=3\\
\hline\hline q=0 & q=2 & q=3 & 
\\\hline \end{array}\:,\label{eq:PRbox3}
\end{equation}
which is a PR box of rank 3, a system whose degree of contextuality
is maximal among all cyclic rank-3 systems in any reasonable theory
of contextuality. 

\section{Why is this happening in CbD?}

What shall one do now with the expectation that (\ref{eq:contextual 1})
should not have been contextual because $\mathcal{R}_{0}$ is informationally
redundant? Its plausibility depends on the definition of contextuality,
because this expectation clearly does not hold for just any property
of the system. Think, e.g., of such properties as the number of connections
in the system, or the product of the random variables within a given
context. To give a more remote but apt analogy, consider the notion
of a rank of a matrix. Adding to a matrix a new column whose entries
are squared values of one of the old columns would generally increment
its rank by 1, even though the new column is informationally redundant. 

It just happens that contextuality is one of such properties. Contextuality
is not about predicting values of random variables in a system. Rather
it is about the compatibility of certain couplings for its connections
with the distributions of its bunches. The answer as to why such compatibility
generally changes as one adds new variables to the system, even if
computable from other variables, has been given in Ref. \cite{DzhCerKuj2017},
and we will present it using our example. The system in (\ref{eq:variant 1 with addition 1})
is noncontextual because it has a coupling with (multi)maximal connections,
 In fact this coupling happens to be unique:
\begin{equation}
\begin{array}{|c|c|c|c||c|}
\hline S & S & \boxempty & 1 & c=1\\
\hline \boxempty & S & S & \boxempty & c=2\\
\hline S & \boxempty & S & -1 & c=3\\
\hline\hline q=1 & q=2 & q=3 & q=4 & 
\\\hline \end{array}\:,
\end{equation}
where $S$ is as in (\ref{eq:coupling 0}). It is easy to see that
the only coupling of $\mathcal{R}_{0}$ compatible with coupling (\ref{eq:coupling 0})
is 
\begin{equation}
\begin{array}{|c||c|}
\hline S & c=1\\
\hline \boxempty & c=2\\
\hline -S & c=3\\
\hline\hline q=0 & 
\\\hline \end{array}\:,
\end{equation}
and this is clearly not a (multi)maximal coupling: $p\left[S=-S\right]=0$,
whereas in the maximal coupling of $\mathcal{R}_{0}$ (since $X$
and $-Z$ are identically distributed) the probability would be 1.
This means that there exists no coupling of the entire system (\ref{eq:contextual 1})
in which all connections are (multi)maximal, and this, by definition,
means that the system is contextual. 

\section{\label{sec:What-if-all}What if all systems are consistently connected?}

The question one can pose now is: how critical is it that in order
to transform the noncontextual system (\ref{eq:variant1 cyclic3})
into the contextual system (\ref{eq:contextual 1}) we have created
a system, (\ref{eq:variant 1 with addition 1}), which is inconsistently
connected? The answer is that it is not critical at all. To see this,
let us modify our example, and replace (\ref{eq:variant 1 with addition 1})
with
\begin{equation}
\begin{array}{|c|c|c|c|c||c|}
\hline X & X & \boxempty & 1 & \boxempty & c=1\\
\hline \boxempty & Y & Y & \boxempty & \boxempty & c=2\\
\hline Z & \boxempty & Z & \boxempty & 1 & c=3\\
\hline\hline q=1 & q=2 & q=3 & q=4 & q=5 & 
\\\hline \end{array}\:.\label{eq:addition 2}
\end{equation}
This system is noncontextual and consistently connected, but we can
still introduce the same connection $\mathcal{R}_{0}$ as before by
means of the following function of the existing connections:
\begin{equation}
\mathcal{R}_{0}=\left\{ \begin{array}{ccc}
\mathcal{R}_{1} & \textnormal{if} & \mathcal{R}_{4}=1\\
\mathcal{-R}_{1} & \textnormal{if} & \mathcal{R}_{5}=1\\
\boxempty & \textnormal{if} & \textnormal{otherwise}
\end{array}\right\} =g\left(\mathcal{R}_{1},\mathcal{R}_{4},\mathcal{R}_{5}\right).
\end{equation}
We end up with essentially the same contextual system as (\ref{eq:contextual 1}),
but one satisfying the condition of consistent connectedness: 
\begin{equation}
\begin{array}{|c|c|c|c|c|c||c|}
\hline X & X & X & \boxempty & 1 & \boxempty & c=1\\
\hline \boxempty & \boxempty & Y & Y & \boxempty & \boxempty & c=2\\
\hline -Z & Z & \boxempty & Z & \boxempty & 1 & c=3\\
\hline\hline q=0 & q=1 & q=2 & q=3 & q=4 &  & 
\\\hline \end{array}\:.
\end{equation}

\section{Strong consistent connectedness}

In the cause of defending the ``core principle'' of informational
redundancy one could point out that the last system, while consistently
connected, is not \emph{strongly} consistently connected \cite{chapter},
because it contains the subsystem 
\begin{equation}
\begin{array}{|c|c||c|}
\hline X & X & c=1\\
\hline -Z & Z & c=3\\
\hline\hline q=0 & q=1 & 
\\\hline \end{array}\:,
\end{equation}
in which the bunches $\left(X,X\right)$ and $\left(-Z,Z\right)$
are not identically distributed. In a strongly consistently connected
system, if two contexts contain variables with the matching contents,
say, 
\begin{equation}
\begin{array}{c}
R_{1}^{c},\ldots,R_{k}^{c}\\
R_{1}^{c'},\ldots,R_{k}^{c'}
\end{array},
\end{equation}
their joint distributions are the same. This constraint, for instance,
is imposed in the sheaf-theoretic approach to contextuality \cite{AbramskyBrandenburger2011}. 

It is a debatable point whether consistent connectedness which is
not strong should also be considered form of disturbance, but we can
avoid this discussion. Examples with all systems involved being strongly
consistently connected can be readily constructed. Consider the following
noncontextual system,
\begin{equation}
\begin{array}{|c|c||c|}
\hline R_{1}^{1} & R_{2}^{1} & c=1\\
\hline \boxempty & R_{2}^{2} & c=2\\
\hline R_{1}^{3} & \boxempty & c=3\\
\hline\hline q=1 & q=2 & 
\\\hline \end{array}=\begin{array}{|c|c||c|}
\hline X & X & c=1\\
\hline \boxempty & Y & c=2\\
\hline Z & \boxempty & c=3\\
\hline\hline q=1 & q=2 & 
\\\hline \end{array}\:,\label{eq:exmapleJanne}
\end{equation}
where $X,Y,Z$ have the same properties as before. Define a new connection
$\mathcal{R}_{0}$ as follows:
\begin{equation}
\mathcal{R}_{0}=\left\{ \begin{array}{ccc}
\mathcal{R}_{2} & \textnormal{if} & \mathcal{R}_{1}=\boxempty\\
\mathcal{-R}_{1} & \textnormal{if} & \mathcal{R}_{2}=\boxempty\\
\boxempty & \textnormal{if} & \textnormal{otherwise}
\end{array}\right\} =\phi\left(\mathcal{R}_{1},\mathcal{R}_{2}\right).
\end{equation}
The new system
\begin{equation}
\begin{array}{|c|c|c||c|}
\hline \boxempty & X & X & c=1\\
\hline Y & \boxempty & Y & c=2\\
\hline -Z & Z & \boxempty & c=3\\
\hline\hline q=0 & q=1 & q=2 & 
\\\hline \end{array}\label{eq:exmapleJanne2}
\end{equation}
is contextual. In fact, it is again a PR box of rank 3, like the one
we had in (\ref{eq:PRbox3}). Both the initial system (\ref{eq:exmapleJanne})
and the resulting system (\ref{eq:exmapleJanne2}) are strongly consistently
connected.

\section{Informationally interchangeable connections}

Returning to the example of Section \ref{sec:What-if-all}, it also
can be modified to involve only strongly consistently connected systems.
To achieve this, however, we need the following modification of the
``core principle'' of informational redundancy. Suppose that we
have two systems, one with connections $\left(\mathcal{R}_{1},\mathcal{R}_{2},\ldots,\mathcal{R}_{k}\right)$
and the other with connections $\left(\mathcal{R}_{0},\mathcal{R}_{2},\ldots,\mathcal{R}_{k}\right)$,
such that we have both
\begin{equation}
\mathcal{R}_{0}=f\left(\mathcal{R}_{1},\mathcal{R}_{2},\ldots,\mathcal{R}_{k}\right)
\end{equation}
and
\begin{equation}
\mathcal{R}_{1}=g\left(\mathcal{R}_{0},\mathcal{R}_{2},\ldots,\mathcal{R}_{k}\right).
\end{equation}
In other words, given the connections $\mathcal{R}_{2},\ldots,\mathcal{R}_{k}$,
the connections $\mathcal{R}_{0}$ and $\mathcal{R}_{1}$ are \emph{informationally
interchangeable}. Then, we suggest, anyone who considers the principle
of informational redundancy intuitively plausible, should also accept
the following principle: if a system with $\left(\mathcal{R}_{1},\mathcal{R}_{2},\ldots,\mathcal{R}_{k}\right)$
is noncontextual, then the system with $\left(\mathcal{R}_{0},\mathcal{R}_{2},\ldots,\mathcal{R}_{k}\right)$
should be noncontextual too. With the alleged principle thus modified,
we can present our counter-example to it by the chain of transformations
in which the systems considered at each step are strongly consistently
connected:
\begin{equation}
\begin{array}{c}
\begin{array}{|c|c|c||c|}
\hline X & X & \boxempty & c=1\\
\hline \boxempty & Y & Y & c=2\\
\hline Z & \boxempty & Z & c=3\\
\hline\hline q=1 & q=2 & q=3 & 
\\\hline \end{array}\\
\downarrow\\
\begin{array}{|c|c|c|c|c||c|}
\hline X & X & \boxempty & 1 & \boxempty & c=1\\
\hline \boxempty & Y & Y & \boxempty & \boxempty & c=2\\
\hline Z & \boxempty & Z & \boxempty & 1 & c=3\\
\hline\hline q=1 & q=2 & q=3 & q=4 & q=5 & 
\\\hline \end{array}\\
\downarrow\\
\begin{array}{|c|c|c|c|c||c|}
\hline X & X & \boxempty & 1 & \boxempty & c=1\\
\hline \boxempty & Y & Y & \boxempty & \boxempty & c=2\\
\hline -Z & \boxempty & Z & \boxempty & 1 & c=3\\
\hline\hline q=0 & q=2 & q=3 & q=4 & q=5 & 
\\\hline \end{array}
\end{array}\:.
\end{equation}
The final transition here is justified by observing that we have both
\begin{equation}
\mathcal{R}_{0}=\left\{ \begin{array}{ccc}
\mathcal{R}_{1} & \textnormal{if} & \mathcal{R}_{4}=1\\
\mathcal{-R}_{1} & \textnormal{if} & \mathcal{R}_{5}=1\\
\boxempty & \textnormal{if} & \textnormal{otherwise}
\end{array}\right\} =g\left(\mathcal{R}_{1},\mathcal{R}_{4},\mathcal{R}_{5}\right),
\end{equation}
and

\begin{equation}
\mathcal{R}_{1}=\left\{ \begin{array}{ccc}
\mathcal{R}_{0} & \textnormal{if} & \mathcal{R}_{4}=1\\
\mathcal{-R}_{0} & \textnormal{if} & \mathcal{R}_{5}=1\\
\boxempty & \textnormal{if} & \textnormal{otherwise}
\end{array}\right\} =h\left(\mathcal{R}_{0},\mathcal{R}_{4},\mathcal{R}_{5}\right).
\end{equation}

\section{A note on indicator connections}

Many similar examples can be readily constructed. Moreover, for a
broad class of systems like those considered in our example one can
formulate a general algorithm by which one can introduce new connections
using the old connections and some auxiliary ones. The latter prominently
include the ``indicator connections'' each consisting of a single
random variable. In our example we conveniently chose indicator variables
as those equal to 1 with probability 1. However, they can be replaced
with arbitrary random variables, not necessarily deterministic ones.
Consider, for example, 

\begin{equation}
\begin{array}{|c|c|c|c|c||c|}
\hline X & X & \boxempty & V & \boxempty & c=1\\
\hline \boxempty & Y & Y & \boxempty & \boxempty & c=2\\
\hline Z & \boxempty & Z & \boxempty & W & c=3\\
\hline\hline q=1 & q=2 & q=3 & q=4 & q=5 & 
\\\hline \end{array},
\end{equation}
where $V$ and $W$ are $\pm1$-variables with $p\left[V=1\right]=p\left[W=1\right]=1/2$,
chosen so that the system remains noncontextual. For instance, one
can introduce $V$ as stochastically independent of $X$, and $W$
as stochastically independent of $Z$. Alternatively, one could put
$V=X$ and $W=Z$. It is easy to check that in either case the system
is noncontextual. Then one can define the new connection $\mathcal{R}_{0}$
in the following way:
\begin{equation}
\mathcal{R}_{0}=\left\{ \begin{array}{ccc}
\mathcal{R}_{1} & \textnormal{if} & \mathcal{R}_{4}=1\textnormal{ or }\mathcal{R}_{4}=-1\\
\mathcal{-R}_{1} & \textnormal{if} & \mathcal{R}_{5}=1\textnormal{ or }\mathcal{R}_{5}=-1\\
\boxempty & \textnormal{if} & \textnormal{otherwise}
\end{array}\right\} =f\left(\mathcal{R}_{1},\mathcal{R}_{4},\mathcal{R}_{5}\right).
\end{equation}
The reason this works is that if content $q$ is not measured in some
context $c$, then the answer to the question
\begin{quote}
\begin{center}
``does the outcome of measuring $q$ in $c$ equal either $1$ or
$-1$?''
\par\end{center}

\end{quote}
must be ``no.'' Moreover, one can replace ``equal to either 1 or
$-1$'' with any numerical relation that is always true about a $\pm1$-variable,
such as $\mathcal{R}_{4}<2$ or $R_{4}\in\mathbb{R}$.

\section{\label{sec:Consistification}Consistification}

If one's goal is to come up with a statement that would contradict
CbD but spare its special case, confined to consistently connected
systems, the only way to do this is to somehow prohibit, in formulations
of functions, access to information about contexts. One simple way
to do this is to only allow functions $\mathcal{R}_{0}=f\left(\mathcal{R}_{1},\ldots\mathcal{R}_{k}\right)$
such that
\begin{equation}
\text{\ensuremath{\mathcal{R}_{0}}}=\boxempty\textnormal{ if }\mathcal{R}_{i}=\boxempty\text{ for some }i=1,\ldots,k.\label{eq:constraint}
\end{equation}
With this constraint, if the old system is consistently connected
and noncontextual, so will be the new system, with $\mathcal{R}_{0}$
added. What happens here is that the combination of (\ref{eq:constraint})
and consistent connectedness amounts to denying contexts any role
in determining the distributions of the random variables and functions
thereof. Such a constraint, of course, is antithetical to any theory
that allows inconsistently connected systems --- because, by definition,
in such systems measurement outcomes are determined by both contents
and contexts. Also, if the claim is that the principle of informational
redundancy is intuitively plausible, it is difficult to see how one
would justify prohibition of the functions considered in the previous
sections of this paper.

Whatever the justification proposed, however, there is an argument
against this or any other attempt to dismiss CbD while preserving
the theory of consistently connected systems. Consider the following
systems:
\begin{equation}
\begin{array}{|c|c|c||c|}
\hline X & X' & \boxempty & c=1\\
\hline \boxempty & Y & Y' & c=2\\
\hline Z' & \boxempty & Z & c=3\\
\hline\hline q=1 & q=2 & q=3 & \textnormal{system }\mathcal{A}
\\\hline \end{array}
\end{equation}
and
\begin{equation}
\begin{array}{|c|c|c|c|c|c||c|}
\hline X & X' & \boxempty & \boxempty & \boxempty & \boxempty & c=1\\
\hline \boxempty & V_{X'} & V_{Y} & \boxempty & \boxempty & \boxempty & c=12\\
\hline \boxempty & \boxempty & Y & Y' & \boxempty & \boxempty & c=2\\
\hline \boxempty & \boxempty & \boxempty & V_{Y'} & V_{Z} & \boxempty & c=23\\
\hline \boxempty & \boxempty & \boxempty & \boxempty & Z & Z' & c=3\\
\hline V_{X} & \boxempty & \boxempty & \boxempty & \boxempty & V_{Z'} & c=31\\
\hline\hline q=1' & q=2'' & q=2' & q=3'' & q=3' & q=1'' & \textnormal{system }\mathcal{B}
\\\hline \end{array}\:.
\end{equation}
System $\mathcal{A}$ is a realization of (\ref{eq:cyclic3}), but
this time $X,X',Y,Y',Z,Z'$ are arbitrary dichotomic variables. That
is, system $\mathcal{A}$ is generally inconsistently connected. System
$\mathcal{B}$ is a \emph{consistification} of system $\mathcal{A}$,
the notion elaborated in Ref. \cite{chapter}, where one can find
a general algorithm for consistifying arbitrary systems.\footnote{This construction was first described in Ref. \cite{Amaral etal.2018},
except for using maximal rather than multimaximal couplings. This
makes no difference for cyclic systems.} In system $\mathcal{B}$, the bunches of the variables in contexts
$c=1,2,3$ are the same as in system $\mathcal{A}$, although their
contents are redefined to make these bunches non-overlapping. There
are also new contexts inserted between the old ones, and filled with
new random variables. Each variable $V_{T}$ has the same distribution
as variable $T$ ($=X,X',Y,\ldots$), and when in the same context,
$V_{T}$ and $V_{U}$ are jointly distributed so that $p\left[V_{T}=V_{U}\right]$
has the maximal possible value (given their individual distributions).
Clearly, system $\mathcal{B}$ is consistently connected, in fact
strongly so. 

Now, it can be shown that systems $\mathcal{A}$ and $\mathcal{B}$
are \emph{contextually equivalent}, in the following sense \cite{chapter}.\emph{
First}, $\mathcal{A}$ is contextual if and only if so is $\mathcal{B}$.
\emph{Second}, if $\mathcal{A}$ is consistently connected, then $\mathcal{A}$
and $\mathcal{B}$ have the same value of \emph{contextual fraction},
as defined in Ref. \cite{Abramskyetal.2017}. \emph{Third}, whether
$\mathcal{A}$ is or is not consistently connected, $\mathcal{A}$
and $\mathcal{B}$ have the same value of any of the CbD measures
of contextuality defined in Ref. \cite{DKC2020}.\footnote{Of these measures, $\mathsf{CNT_{0}}$ in $\mathcal{A}$ equals $\textnormal{CNT}_{2}$
in $\mathcal{B}$ because they are computed by the same linear programming
algorithm. Since both systems are cyclic ($\mathcal{B}$ being cyclic
of rank 6), $\mathsf{CNT_{0}=CNT_{1}=CNT_{2}}$ for either of them
\cite{DKC2020}.} 

The point of this demonstration is that all systems in CbD can always
be redefined in terms of their consistifications, i.e., using strongly
consistently connected systems only. For instance, an empirical situation
that could be described by system $\mathcal{A}$ above can instead
be required to be described by system $\mathcal{B}$. If one did this
systematically, the resulting theory (let us call it CbD{*}) would
become more cumbersome but would not change in any significant aspect.
It is difficult to see how one can find problems with CbD while accepting
CbD{*} if the two formulations are essentially equivalent. 

\section{Concluding remarks}

Rather than continue to elaborate and multiply examples, let us state
the main conclusion of this paper, for which a single example is all
one needs. In a theory of contextuality based on multimaximally connected
couplings, adding an informationally redundant connection or exchanging
it for an informationally equivalent one may change a noncontextual
system into a contextual one. This is also true for theories confined
to consistently connected systems, and even strongly consistently
connected systems. Therefore this fact is inherent to the very notion
of contextuality.

The discussion of the connections and their functions in this paper
has an unexpected benefit: a better understanding of the role of non-measurements
in contextuality analysis. We now know that non-measurements in systems
of random variables should be treated as elements of these systems
on a par with the measurements represented by random variables. The
information provided by a non-measurement, that a given content is
not measured in a given context, is accessible, and it can be utilized,
in particular, in defining functions mapping connections into connections.
In CbD, non-measurements can be replaced with deterministic values,
as shown in (\ref{eq:CbDdet}). They can also be viewed as random
variables that are equal to the special value $\boxempty$ with probability
1 (see Footnote \ref{fn:We-need-to}). For instance, our system (\ref{eq:cyclic3})
is represented by 
\begin{equation}
\begin{array}{|c|c|c||c|}
\hline R_{1}^{1} & R_{2}^{1} & R_{3}^{1}\equiv\boxempty & c=1\\
\hline R_{1}^{2}\equiv\boxempty & R_{2}^{2} & R_{3}^{2} & c=2\\
\hline R_{1}^{3} & R_{2}^{3}\equiv\boxempty & R_{3}^{3} & c=3\\
\hline\hline q=1 & q=2 & q=3 & 
\\\hline \end{array}\:,
\end{equation}
in which every connection contains two distinct distributions. In
this sense, every system of interest is inconsistently connected,
even if strongly consistently connected in the traditional sense.
A strongly consistently connected system without non-measurements
is necessarily trivial, with all contexts containing one and the same
joint distribution.

Moreover, the inclusion of non-measurements as legitimate values of
random variables allows one to extend the notion of a system of random
variables to include situations in which a content within a context
may sometimes be measured and sometimes not. Consider, for instance,
an experiment in which the dichotomous variables $R_{1}^{1}$ and
$R_{2}^{1}$ in context $c=1$ are jointly assigned values ($\pm1$)
only if the detectors of the corresponding properties ``click''
within a short temporal window. In a real experiment, it occasionally
happens that the two clicks are separated by a larger interval, or
one of the two detectors does not click at all. Such experimental
results are necessarily excluded from analysis. With the new point
of view, however, if a detector does not click or clicks too late,
we simply set the corresponding variable equal to $\boxempty$. This
means that $R_{1}^{1}$ and $R_{2}^{1}$ (and, by analogy, all other
variables in the system) are now trichotomous, with the possible values
$\left\{ 1,-1,\boxempty\right\} $ occurring, generally, with non-zero
probabilities. This might potentially lead to more comprehensive contextuality
analysis, more veridically combining quantum properties with those
of the measuring procedures. 

\paragraph{Acknowledgements}

We are indebted to V\'ictor H. Cervantes for critically reading and
commenting on the manuscript, as well as clarifying for us some aspects
of consistified systems. We are grateful to Matthew Jones and Alisson
Tezzin for kindly communicating to us their abstract and answering
questions that clarified its meaning.

\end{document}